\documentclass[preprint,showpacs,preprintnumbers]{revtex4}

\usepackage{graphicx}
\usepackage{dcolumn}
\usepackage{bm}

\newcommand{\be}{\begin{equation}}
\newcommand{\ee}{\end{equation}}
\newcommand{\bea}{\begin{eqnarray}}
\newcommand{\eea}{\end{eqnarray}}
\newcommand{\mev}{\rm MeV}

\begin{document}

\bibliographystyle{apsrev}

\preprint{UAB-FT-519}

\title{Primordial Helium Production in a Charged Universe} 

\author{Eduard Mass{\'o}} 
\email[]{masso@ifae.es}
\author{Francesc Rota}
\email[]{rota@ifae.es}

\affiliation{Grup de F{\'\i}sica Te{\`o}rica and Institut 
de F{\'\i}sica d'Altes
Energies\\Universitat Aut{\`o}noma de Barcelona\\ 
08193 Bellaterra, Barcelona, Spain}


\date{\today}

\begin{abstract}
We use the constraints arising from primordial nucleosynthesis to bound a putative electric charge density $|e|n_q$ of the universe. We find $|n_q/n_\gamma| \lesssim 10^{-43}$, four orders of magnitude more stringent than previous limits. We also work out the bounds on $n_q$ in models with a photon mass, that allows to have a charge density without large-scale electric fields.
\end{abstract}

\maketitle


One of the fundamental parameters of the universe is its electric charge. It is usually assumed to be exactly vanishing, but it is of course desirable to have observational evidences of the hypothesis of a neutral universe. In this paper we will discuss constraints on $n_q$, defined in such a way that $|e|n_q$ is the electric charge density of the universe ($e$ is the electron charge). It is convenient to normalize this quantity to the photon number density, so we define the parameter
\be
\eta_q \equiv \frac{n_q}{n_\gamma}
\ee 
with  
\be
n_\gamma = \frac{2\,\xi(3)}{\pi^2}\,T^3
\label{ngamma}
\ee
where $T$ is the temperature, and $\xi(3) \simeq 1.20$.

Speculations about a possible electric charge of the universe go back to the work of Lyttleton and Bondi \cite{lyttleton}, who showed that an excess charge could account for the expansion of the universe. They assumed that the excess charge was due to a tiny charge imbalance among the proton and the electron, $q_p + q_e \neq 0$. Laboratory limits \cite{dylla,Groom:2000in} on this imbalance rule out the assumption of Lyttleton and Bondi. Their argument can be somehow inverted \cite{Orito:1985cf}: demanding that the gravitational attraction among cosmological objects is larger than the electromagnetic repulsion leads to the bound
\be
|\eta_q| \lesssim 10^{-28}
\label{etaq}
\ee 

More recent speculations involving an excess charge are exotic suggestions of violation of electric charge conservation. Grand unified theories imply violations of baryon and lepton numbers, but conservation of electric charge is usually believed to be exact as a result of gauge invariance. Nonetheless one has the possibility of a (necessary small) breakdown of electric charge conservation at high energies \cite{voloshinetal}. Also, we may have an excess charge in theories with extra dimensions, first introduced by Kaluza and Klein \cite{kaluzaklein} and subject to a recent strong revival \cite{extradim}. The extra dimensions should be compactified to a small size, since our world appears four dimensional. To probe them one can consider the universe in its first instants; the energy is so high that one has a truly higher-than-four dimensional space. Subsequently, the universe evolves and cools down but some electric charge could remain in the ordinary space. Some authors have built classes of models where there is charge leakage into infinite extra dimensions \cite{dubovsky}.

It is therefore interesting to examine the implications of a charge density $n_q$ and to find constraints on it. Orito and Yoshimura \cite{Orito:1985cf} examined some of these implications. They considered the anisotropies generated by the large-scale electric field that would be produced by a net charge density. Their best bound is obtained using observational limits on cosmic ray anisotropies:
\be
|\eta_q| \lesssim 10^{-39}
\label{orito}
\ee

In this letter we show that a much more stringent bound can be obtained by considering the primordial nucleosynthesis period of the early universe and the change in the primordial helium production that would be originated by a charge density present at the nucleosynthesis era.

The physics involved in the primordial nucleosynthesis period is well understood and the theoretical predictions of the primordial yields of light elements are robust \cite{Sarkar:1996dd,Olive:2000ij}. The agreement with observation is considered one of the pillars of modern cosmology. For our purposes, it is useful to recall here the simple arguments that allow to understand the main physical features of helium production \cite{Sarkar:1996dd,Kolb:1990vq}. In the early universe, at time $t \ll 1\,$s and temperature $T \gg 1\, \mev$, neutrons and protons are in kinetic and chemical equilibrium due to weak interactions. At a lower $T \approx T_f$, the rate of these interactions becomes less than the expansion rate of the universe and they go out of equilibrium. Then, the relative neutron-to-proton density is given approximately by
\be
\left(\frac{n}{p}\right)_f \simeq e^{-Q/T_f} \approx 0.24 
\label{npratio}
\ee
where
\be
Q = m_n - m_p = 1.29 ~\mev
\label{Q}
\ee
We have taken $T_f \approx 0.9 \,\mev$. After freeze-out, there is a decrease in the neutron number due to $\beta$-decay,
\be
n \,\rightarrow\, p\,\,e^-\,\bar{\nu}
\label{betadecay}
\ee
until the time $t_{He}$ where all the helium is produced. Then
\be
\left(\frac{n}{p}\right)_{He} \simeq e^{-t_{He}/\tau_n}\,\left(\frac{n}{p}\right)_f \approx 0.15 
\label{npratio2}
\ee
Here we have introduced the neutron lifetime \cite{Groom:2000in} $\tau_n = 887\, $s and we have taken $t_{He} \approx 400 \,$s.

Since nearly all neutrons are processed into helium, the expected mass fraction $Y$ of $^4$He is 
\be
Y \simeq \frac{2\,(n/p)_{He}}{1 + (n/p)_{He}} \approx 0.26
\label{YHe}
\ee
The actual theoretical prediction for $Y$ has to be obtained by a numerical code \cite{Wagoner,Kawano} that solves the relevant set of ordinary differential equations, and indeed gives a result not far from (\ref{YHe}).

Primordial nucleosynthesis is known to be a useful tool to constrain non-standard physics \cite{Sarkar:1996dd}. We now apply it to the issue of a charged universe. A charged particle in such universe has a potential $V$ (with the condition $V=0$ when $n_q=0$) and thus it has a momentum $p$ related to its total energy $E$ by
\be
(E-V)^2=p^2+m^2
\label{E-V}
\ee
With the introduction of an effective mass $m^*$, the dispersion relation can be written as 
\be
E^2=p^2+(m^*)^2
\ee
We now assume $V\ll E$ and that the charged particle is non-relativistic. We get 
\be
m^* \simeq m + V
\label{meff}
\ee
A similar phenomenon is at the basis of the MSW effect \cite{MSW}; eqs\,(\ref{E-V}-\ref{meff}) in the relativistic case are discussed in \cite{bethe}.

The horizon distance $a_H$ is defined as the propagation time of light since $t=0$, and then for $r>a_H$ the charge density could not interact with the charged particle. Thus, to the potential $V$ contributes all the charge inside a sphere of radius $a_H$
\be
V = \pm \int_0^{a_H}\,(dr\,4\pi\,r^2)\,\frac{\alpha\,n_q}{r}
\label{V}
\ee

To summarize, a net charge density $|e|n_q$ induces effective masses for the charged particles present in the universe at that time
\bea
m_p ~ \rightarrow ~ m_p^* &=& m_p + \delta m \nonumber \\
m_e ~ \rightarrow ~ m_e^* &=& m_e - \delta m
\label{mstar}
\eea
(positrons are affected with the opposite sign than electrons), where from (\ref{meff}) and (\ref{V}) 
\be
\delta m = 2\pi\,\alpha\, a_H^2\, n_\gamma\, \eta_q
\label{deltam}
\ee
As we are in a radiation dominated universe we will put $a_H = 2\,t$.

A non-vanishing $\delta m$ alters the predictions of primordial nucleosynthesis for the helium yield, and this leads to a bound on $\delta m$ and hence on $\eta_q$. Before presenting our numerical results let us show that we can get an approximate expression for the change in $Y$ due to the $\delta m$ shift, using the same type of simple analysis that we used in our discussion of the standard $Y$, from eq.~(\ref{npratio}) to (\ref{YHe}). We work at first order in $\delta m$.

There are two main sources of change in $Y$. The first is through the dependence of $Y$ in eq.~(\ref{npratio}) on $Q$ at freeze-out,
\be
\frac{\delta Y}{Y} \simeq \frac{1}{1+(n/p)_f}\,\frac{\delta m}{T_f} \approx \frac{\delta m}{\mev}
\label{deltaY1}
\ee
where we have introduced the numerical values for $T_f$ and $(n/p)_f$.

The second is due to the dependence of $\tau_n$, the lifetime of process (\ref{betadecay}), on the proton and electron masses. When decaying into a proton of mass $m_p^*$ and an electron of mass $m_e^*$, the neutron lifetime (for zero temperature) depends on masses as
\bea
\frac{1}{\tau_n^*} &\propto& m_e^{*\,5}\, \lambda(q^*) \\
\lambda(q^*) &=& \int_1^{q^*}\,dx\,x\,(x-q^*)^2\,\sqrt{x^2-1} \nonumber \\
&=& \frac{1}{60}\,\sqrt{q^*-1}\,(2\,q^{*4}-9\,q^{*2}-8) \nonumber \\
&& + \frac{1}{4}\,q^*\,\ln(\sqrt{q^{*2}-1}+q^*)
\eea
with $q^* = (m_n-m_p^*)/m_e^*$

After some algebra one finds the change in $\tau_n$
\be
\frac{\delta\tau_n}{\tau_n} \simeq \left[5-(q-1)\left(\frac{1}{\lambda}\,\frac{d\lambda}{dq^*}\right)_{q^*=q} \right] \frac{\delta m}{m_e}
\label{deltatau}
\ee
with $q = (m_n-m_p)/m_e$. From (\ref{npratio2}) it is easy to see that a shift $\delta \tau_n$ induces a change $\delta Y$
\be
\frac{\delta Y}{Y} \simeq \frac{1}{1+(n/p)_{He}}\,\frac{t_{He}}{\tau_n^2}\,\delta\tau_n
\label{deltaY2}
\ee
Introducing the numerical values for the parameters that appear in (\ref{deltatau}) and (\ref{deltaY2}), we end up with
\be
\frac{\delta Y}{Y} \approx \frac{\delta m}{\mev}
\label{deltaY3}
\ee

To find the expressions (\ref{deltaY1}) and (\ref{deltaY3}) we have made the assumption that $\delta m$ is independent of time, which is not true, since $a_H=2t$ and $n_{\gamma}\sim T^3 \sim t^{-3/2}$ in (\ref{deltam}). Still, we are able to find the right order of magnitude for our bound on $\eta_q$ if we proceed as follows. Since (\ref{deltaY1}) is approximately valid at freeze-out, we evaluate $\delta m$ in (\ref{deltam}) using $t\approx 1\,$s, $T\approx 1\,\mev$ and get
\be
\delta m \approx 10^{41}\,\eta_q~\mev
\label{deltam_1}
\ee
Similarly, (\ref{deltaY3}) is approximately valid during the period of neutron decay. Now we evaluate $\delta m$ in (\ref{deltam}) with typical values for this period, $t\approx 100\, $s, $T\approx 0.1\,\mev$, and get
\be
\delta m \approx 10^{42}\,\eta_q~\mev
\label{deltam_2}
\ee

Comparing (\ref{deltam_1}) and (\ref{deltam_2}), we see that it is the change in $Y$ due to $\delta\tau_n$, eq.~(\ref{deltam_2}), that will dominate the whole effect. The origin of this dominance is that, as we see from (\ref{deltam}), $\delta m \sim t^2\,T^3 \sim t^{1/2}$, i.e., $\delta m$ increases with time. To estimate the order of magnitude of the bound on $\eta_q$, we put (\ref{deltam_2}) in (\ref{deltaY3}) and allow, for instance, a $10\%$ change in $Y$. We get $|\eta_q|\lesssim 10^{-43}$

To find a precise bound, we have implemented the modification (\ref{mstar}) and (\ref{deltam}) in Kawano's version \cite{Kawano} of the primordial nucleosynthesis code of Wagoner \cite{Wagoner} to obtain the effect of a density $n_q$ on the helium abundance. The abundance depends on the laboratory neutron lifetime $\tau_n$, the number of neutrinos $N_\nu$, and the baryon-to-photon ratio $\eta_B$. We use $\tau_n = 887 \pm 2$ s \cite{Groom:2000in}, $N_\nu = 3$ \cite{Groom:2000in}, and for $\eta_B$ the value extracted from CMB anisotropy measurements \cite{CMB}. Unfortunately, this value is still not completely settled but the hope is that in the future it will be with more refined experiments. For the time being, we  take
\be
10^{-10} \le \eta_B \le 10^{-9}
\label{eta_b}
\ee
a generous range that embraces the observations \cite{CMB}.

\begin{figure}[htb]
\begin{center}
\includegraphics[width=5cm, height=8cm, angle=-90]{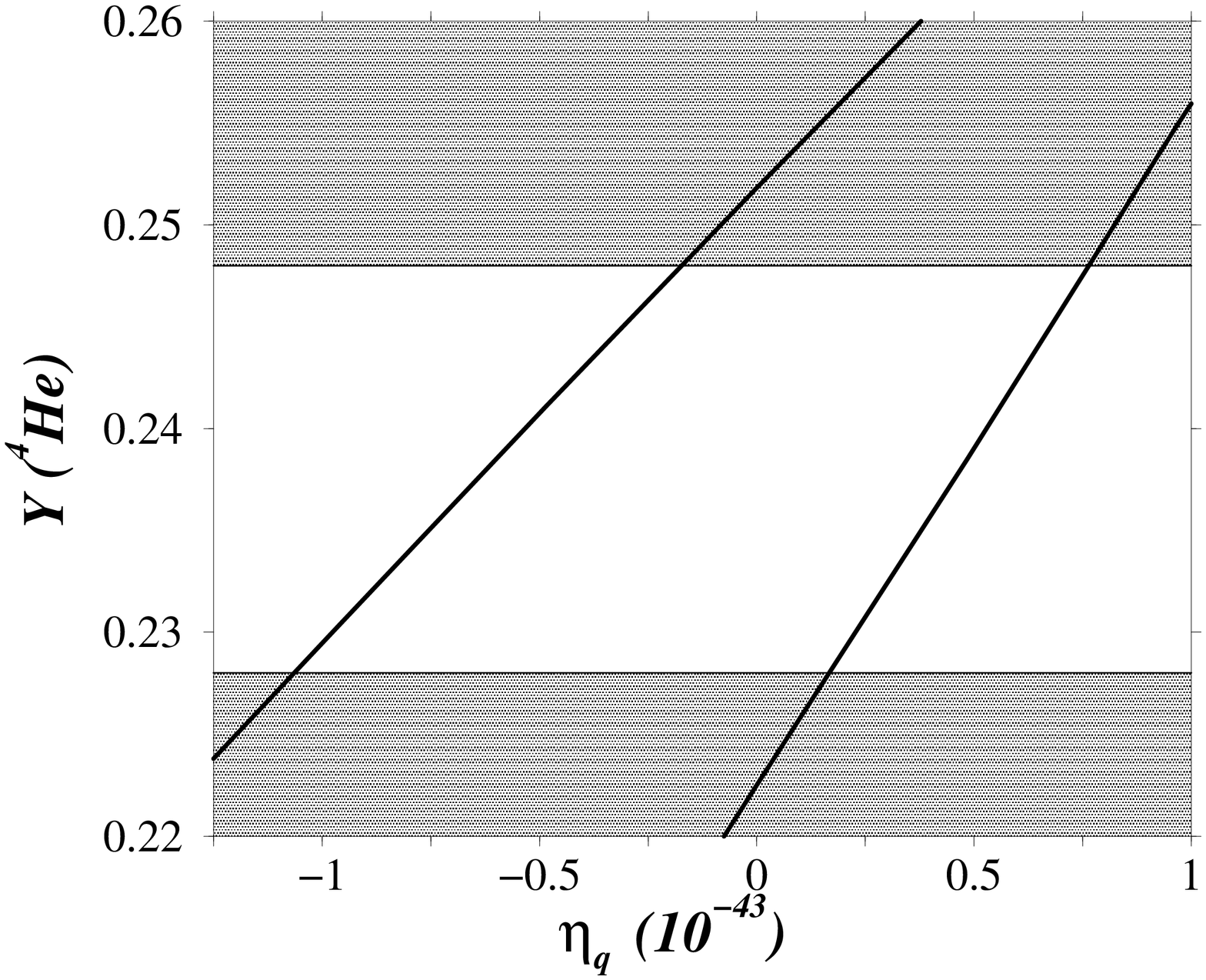}
\end{center}
\caption{\label{fig_1} Predicted value of $Y$ as a function of $\eta_q = n_q/n_\gamma$, for $\eta_B=10^{-9}$ (left-upper line) and for $\eta_B=10^{-10}$ (right-lower line). The two horizontal lines are the observational limits (\ref{Y_obs}). The allowed range of $Y$ is not shadowed.} 
\end{figure}

The prediction for $Y$ as a function of $\eta_q$ is shown in Fig.\ref{fig_1}, for the two extremes of the range (\ref{eta_b}). (The experimental error in $\tau_n$ introduces a negligible change in our prediction.) The linear dependence is due to the expected dominance of the first-order in $\eta_q$. Our bound is stringent because $Y$ is measured with relatively small error. The observational status is discussed in \cite{Sarkar:1996dd,Olive:2000ij}. We adopt the range
\be
0.228 \le Y \le 0.248
\label{Y_obs}
\ee
that the authors of \cite{Olive:2000ij} claim is ``$95 \%$ C.L.''. As we see from Fig.\ref{fig_1}, combining the experimental range (\ref{Y_obs}) with the predictions for $Y$ in the range (\ref{eta_b}) constrains the charge density of the universe. The bound is
\be
-1.1\times10^{-43} \le \eta_q \le 0.8\times10^{-43}
\label{bound}
\ee
We notice that it is four orders of magnitude better than (\ref{orito}), and that it is not far from our estimate using (\ref{deltam_2}).

A non-vanishing $\eta_q$ would also affect the yields of the other light elements D, $^3$He, and $^7$Li. However, since these yields are not as well measured as $^4$He, taking them into account could not significantly improve our bound (\ref{bound}).

Even if an hypothetical charge density of the universe $n_q$ has to be so tiny, one may worry about the induced large-scale electric fields that appear when $n_q \neq 0$, no matter how small. In fact, there are models having $n_q \neq 0$ but no large-scale fields. This can be achieved by endowing the photon with a small mass $m_\gamma$ \cite{Barnes_Barry} and thus having an electromagnetic interaction of finite range $\lambda$. Bounds, like (\ref{orito}), that are based on the effects of a large-scale electric field are very much weakened in this kind of models (to get (\ref{orito}) the authors of \cite{Orito:1985cf} consider a scale $\approx 500$ pc for the electric field).

Let us show that our bound (\ref{bound}), when $m_\gamma \neq 0$, is only modified by one order of magnitude. First, we notice that the mass shift $\delta m \sim t^{1/2}$ shown in (\ref{deltam}) is valid when $a_H = 2\,t < \lambda$. However, for later times, $\delta m$ gets contribution only up to a radius $\lambda$,
\be
\delta m = \frac{4\,\alpha}{\pi}\,\xi(3)\,\eta_q\,\lambda\,T^3 \sim t^{-3/2} 
\label{deltam_3}
\ee
It follows that $\delta m$ increases until $t = \lambda /2$ and afterwards it decreases.

Since helium production finishes when $t \approx t_{He}$, it is clear that our bound (\ref{bound}) is still valid for $\lambda > t_{He}$. For smaller $\lambda$, we expect to find bounds on $n_q$ that are less severe. Obviously, for $\lambda \rightarrow 0$ the bound would disappear. However, $\lambda$ is subject to the experimental constraint
\be
\lambda =  \frac{1}{m_\gamma} \geq 10^9\, \rm m
\label{lambda}
\ee
coming from studies on torques on a toroid balance \cite{Lakes:1998mi}. Similar limits are obtained from measurements of the Jovian magnetic field \cite{Davis:1975mn}. We notice that the lower limit $\lambda = 10^9\,$m is about the same as $t_f \approx 1\,\rm s = 3\times 10^9\,$m. Thus, to estimate the bound on $|\eta_q|$ we can use (\ref{deltam_1}) since it is valid for $t\approx 1\,$s. We expect to get $|\eta_q| \lesssim 10^{-42}$, so that in models with a photon mass we expect to have a bound that is only about one order of magnitude worse than (\ref{bound}). 

\begin{figure}[htb]
\begin{center}
\includegraphics[width=5cm, height=8.5cm, angle=-90]{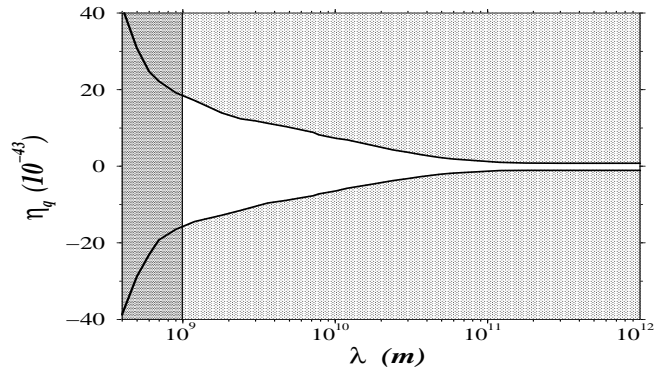}
\end{center}
\caption{\label{fig_2} Upper and lower bounds on $\eta_q$ when electromagnetic interactions have a range $\lambda = 1/m_\gamma$. The experimental limit on $\lambda$ (\ref{lambda}) is displayed as a vertical line. Non-excluded values are in the non-shadowed part of the figure} 
\end{figure}

We have modified Kawano's code \cite{Kawano} with the mass shift $\delta m$ shown in (\ref{deltam_3}) to find the precise change in $Y$. By demanding that the predicted $Y$ is in the experimental interval (\ref{Y_obs}), and allowing for $\eta_B$ in the range (\ref{eta_b}), we are able to limit $\eta_q$. In Fig.\ref{fig_2}, we show our bound as a function of $\lambda$ and also the experimental limit (\ref{lambda}). We confirm that for $\lambda \gtrsim 10^{11}\,$m we get our previous limit (\ref{bound}) and we also see that the constraints relax for smaller $\lambda$. We can put a limit on $\eta_q$ in models with a photon mass, which clearly is the one for $\lambda = 10^9\,$m. It reads
\be
-1.6 \times 10^{-42}\le \eta_q \le 1.8 \times 10^{-42}
\ee 

\begin{acknowledgments}
Discussions with F.~Ferrer, J.~Garriga, J.~A.~Grifols, S.~Sarkar, G.~Senjanovi{\'c} and R.~Toldr{\`a} are greatfully acknowledged. Work partially supported by the CICYT Research Project AEN99-0766, and by the EU network on Supersymmetry and the Early Universe (HPRN-CT-2000-00152)
\end{acknowledgments}




\newpage

\end{document}